\documentclass[aps,prl,twocolumn]{revtex4}
\usepackage{epsfig,psfig}
\begin{document}
%\twocolumn[\hsize\textwidth\columnwidth\hsize\csname@twocolumnfalse%
%\endcsname
\title{\bf Compensation, interstitial defects and ferromagnetism in diluted semiconductors}

\author{Georges ~Bouzerar$^{1,2}$\footnote{email:georges.bouzerar@grenoble.cnrs.fr and bouzerar@ill.fr}, Timothy Ziman$^{2}$\footnote{and CNRS 
email:ziman@ill.fr} and Josef ~Kudrnovsk\'y'$^{2,3}$\footnote{email:kudrnov@fzu.cz}}

\address{ 
$^{1}$ Laboratoire Louis N\'eel, 25 avenue des Martyrs, BP 166 38042 Grenoble Cedex 09
France.\\   
$^{2}$Institut Laue Langevin 
BP 156
38042 Grenoble 
France.\\            
$^{3}$ Institute of Physics, Academy of Sciences of the Czech Republic,Na Solvance 2,CZ-182 21 Prague 8, Czech Republic }
\date{\today}

\begin{abstract}
\parbox{14cm}{\rm}
\medskip
We present a quantitative theory for ferromagnetism in diluted III-V ferromagnetic semi-conductors
in the presence of the two  types of defects
commonly supposed to be responsible for compensation: As anti-sites and Mn interstitials.
In each case we reduce the description to that of an effective random
Heisenberg model with exchange integrals between active magnetic impurities provided by
 ab initio calculation.  The effective magnetic Hamiltonian is then
solved by a  semi-analytical method (locally self-consistent RPA), where disorder is treated exactly.
Measured Curie temperatures are shown to be inconsistent with the hypothesis that  As anti-sites provide the dominant mechanism for compensation. 
In contrast, if we assume that Mn interstitials are the main source for compensation,
we obtain a very good agreement between the calculated 
Curie temperature and the measured values, in both as-grown and annealed samples. 
%We argue that this work provides a clear evidence for the importance and role of
% Mn interstitials in  non-optimally annealed samples of GaMnAs.
%We also  predict  the variation of the Curie temperature as a function of
%hole density in well-annealed samples.
\end{abstract} 

\maketitle

%]
\narrowtext

Diluted Magnetic Semiconductors are  materials where the interplay of transport
and magnetic properties open the perspectives of exciting applications.
The III-V semiconductors are particularly promising since a  low
concentration of magnetic dopants can give relatively high Curie temperatures for ferromagnetism\cite{Ohno,Edmonds}.
In these materials it is found that the Curie temperatures  depend strongly
on methods of preparation and sample history: for the same nominal concentration of magnetic ions
the Curie temperature may vary by large factors. Systematic studies show that different annealing
treatments display a clear  correlation between the 
Curie temperature and the conductivity. This indicates that
the  process of  magnetic     doping
is more complex than a straight substitution  (Mn(Ga)) of (formally)    Ga$^{3+}$ sites by  Mn$^{2+}$ atoms,  providing
a localized magnetic moment and an itinerant hole. In fact the original samples are ``compensated'', i.e. 
the density of holes measured by transport is {\it lower} than the concentration
of magnetic ions due to additional donor impurities especially
in the samples as grown by Molecular Beam Epitaxy (MBE).
%Transport measurements show that this simple picture of substitution is insufficient, especially
%in the samples as grown by Molecular Beam Epitaxy (MBE). 
The increase in T$_C$ after annealing is then interpreted
as removal of the defects, resulting in an increase in the hole  concentration which mediates the magnetic exchange. This leaves 
obscure  the precise form  of the compensating defect,
and does not provide a quantitative theory relating the Mn$^{2+}$ concentration, the hole density and density of compensating defects to  ferromagnetism.
\par
There are two probable candidates for compensation:
both Arsenic antisites As$_{{\rm Ga}}$ (i.e. As atoms on sites of the Ga sub-lattice) and Mn interstitials Mn$_I$ have long been known to be double donors. 
The two forms of defects differ in an important
way: for each  As$_{{\rm Ga}}$ there are two holes removed, i.e. only the carrier density is
changed,  while each interstitial, in addition, introduces a magnetic moment, changing the number of magnetically active ions. Microscopic calculations indicate that 
the Mn$_I$ are preferentially situated on interstitial sites adjacent to 
occupied  Mn(Ga) and that the coupling between interstitials and 
the adjacent moment is essentially given by antiferromagnetic superexchange coupling
( J $\approx $ -320 ~K) \cite{Masek}.
In fact   there are two inequivalent interstitial positions,
the  Mn atom can be  located inside the tetrahedron formed by either four Ga $T({\rm Ga}_{4})$  or else four As  $T({\rm As}_{4})$. 
We  shall make no distinction between the two possible positions
sites, since in either case the  AF exchange with Mn(Ga) is strongly antiferromagnetic,
and we refer to Mn$_{I}$ as the sum of the two.

The immediate question is, what is the proportion of the two defects:
interstitials Mn$_I$ and As$_{{\rm Ga}}$ in  the ferromagnetic
samples? A conclusion of this paper is  that the observed Curie temperatures
in samples at different stages of annealing,
% and therefore
%with different carrier densities, 
can {\it only} be explained  
assuming that  
interstitial defects dominate compensation. Such a dominance 
 agrees  with  Wolos et al. \cite{Wolosetal},
who estimated, from the strength of optical transitions,
a relatively small (fewer than 10\% of the total
Manganese atoms) number of antisites As$_{{\rm Ga}}$, fewer than 10\% of the total
Manganese atoms, and, 
from  Electron Paramagnetic Resonance, a much larger number 
of other compensating defects.  Similarly Wang et al \cite{GallagherEdmonds} 
 showed that the saturated magnetization at low temperatures was  
consistent with the elimination of {\it two} magnetic moments with each impurity.
Furthermore polarised neutrons reflectometry \cite{Kirby} and Auger spectroscopy and resistivity measurements 
\cite{Edmonds}  showed that the annealing process corresponds to redistribution of 
Mn sites and the increase of the magnetisation far from the surface.
We emphasise that clear proof of the r\^ole
of interstitials is still necessary, as other
techniques, by Transmission Electron Micrography\cite{Glasetal},
 or by Infrared Absorption and Positron Annihilation Spectroscopy  \cite{Tuomistoetal} suggested a much higher concentration of anti-sites.
The new element we are bringing here, is a  {\it quantitative} theory
for the Curie temperature, which as we shall explain below,
is much more accurate than Zener mean-field theory.
We note that in ref  \cite{Bouzerar-epl} we anticipated the fact that the changes in the carrier density due
solely to anti-sites were {\it insufficient} to explain the reduction
of T$_C$.
% as the curve
%decreases slowly above a sharp threshold.
\par
Recently, by combining first principles calculations and an semi-analytical approach, we were able to  
provide an excellent agreement between the  calculated Curie temperatures and those measured  in
optimally doped semiconductors \cite{Bouzerar-epl}.
In the first step of this method, we  derive the exchange integrals between magnetic impurities 
using the Local Density Approximation (TB-LMTO) and  magnetic force theorem \cite{Josephetal}
providing an effective classical random Heisenberg Hamiltonian.
Note that 
%the one particle Green's functions of the itinerant carriers used for the calculations of
the calculated exchange integrals, include the effect of disorder within a Coherent Potential Approximation (CPA) for electronic motion.
In the second step, we treat the random effective Heisenberg model within an approach
where thermal fluctuations are treated within a local random phase appoximation (RPA), whilst the disorder is treated '{\it exactly}', i.e. 
the magnetic properties are calculated for individual  configuration of disorder generated by sampling techniques.
This theory is an extension of ref.\cite{Georges2} where disorder in the {\it effective} Hamiltonian was treated by a form of CPA.
An attractive feature  of an ``exact''  treatment of disorder is that it    allows us, for example, to study the effect of correlations in 
the disorder \cite{Bouzerar-apl}, a question of importance in interpreting experiments on  high Curie-temperature samples grown using the OMVPE technique (Organo-Metallic Vapor Phase Epitaxy)  
\cite{Wessels}.
We attribute the  success of our approach  to (i) the realistic calculations of the exchange integrals
(ii) to a proper treatment of the thermal fluctuations and disorder of the effective Heisenberg  Hamiltonian. This second point is confirmed by consistency with Monte-Carlo simulations \cite{Joseph-MC}.
%The importance of the second point is illustrated by comparing to ref.\cite{Josephetal} \cite{Satoetal}, 
%where the effective Heisenberg model was calculated in the same way, but was then solved within a Mean Field-Virtual Crystal Approximation. This lead to exaggerated Curie temperatures.  
%An alternative has been to make model calculations  which start from 
%an effective Hamiltonian using simplified, but realistic, band structures and
%treat  the coupling between carriers and magnetic impurities perturbatively\cite{Dietl}. We maintain that the shortcoming of such an approach has been again in  solving the resultant effective Heisenberg
%Hamiltonian rather than in the effective parameters. 
\par
 We first analyze the dependence of the Curie temperature $T_{C}$ with the density of As anti-sites
$y_{\bar{{\rm As}}}$. As in our preliminary study of this issue\cite{Bouzerar-epl}, which 
was restricted to a single nominal concentration of Mn, we do this by introducing 
in the ab-initio stage of our calculation a concentration of anti-sites, which influences
the calculated exchange couplings. This calculation also allows us to verify, within
LDA  that there is indeed the compensation expected.
In Fig. 1 we plot, for different concentration $x_{{\rm Mn}}$ of Mn, the variation of the 
calculated $T_{C}$ in $({\rm Ga}_{1-x_{{\rm Mn}}-y_{\bar{{\rm As}}}} {\rm Mn}_{x_{{\rm Mn}}} \bar{{\rm As}}_{y_{\bar{{\rm As}}}}){\rm As}$ 
as a function of $\gamma=n_{h}/x_{{\rm Mn}}$. Since each As anti-site is a double donor, the  carrier density is  $n_{h}=x_{{\rm Mn}}-2y_{\bar{{\rm As}}}$.
 We observe that above a critical value  $\gamma_{c}(x_{{\rm Mn}})$ , $T_{C}$ is  weakly sensitive to As-antisites, this is particularly clear for $x_{{\rm Mn}}=0.03$ and 0.05.
For  7\% Mn, for example, we also observe that for $\gamma < 0.50$ ferromagnetism is unstable. The reason for this is that as the 
density decreases,  the nearest neighbour exchange integral  becomes 
increasingly dominated by the (antiferromagnetic) superexchange contribution, leading to frustration.
Note also that this behaviour with hole density is incompatible with Zener mean-field theory which predicts that  $T_{C} \propto n_{h}^{1/3}$. Let us now  discuss the compatibility of these results with the 
assumption that As anti-sites dominate the mechanism
of compensation. We do this by comparing  samples with approximately fixed total density of Mn impurities but exhibiting
large variation in their Curie temperatures.
%In Table 1 we show the variation of $T_{C}$ for a fixed total density of Mn $x_{\rm Mn} \approx 0.067$ and for both as-grown and annealed samples under different conditions. For each sample both the Curie temperature and the density of carrier is provided by transport measurements.

\begin{figure}[tbp]
\centerline{
\psfig{file=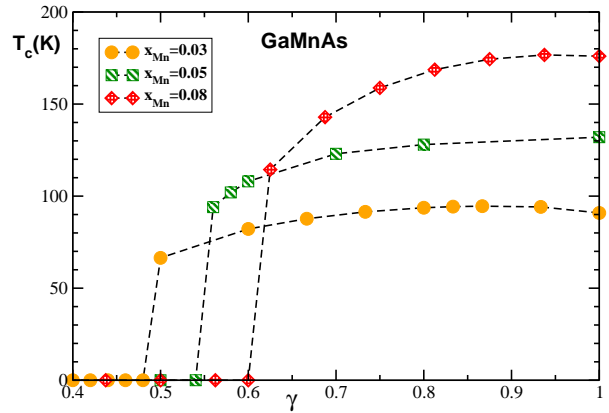,width=7cm,angle=-90}}
\caption{Effect of As anti-sites on $T_C$ in $({\rm Ga}_{1-x_{{\rm Mn}}-y_{\bar{\rm As}}}{\rm Mn}_{x_{{\rm Mn}}}{\rm As}
_{y_{\bar{\rm As}}}){\rm As}$ for different Mn concentation. The carrier density is $n_{h}=x_{{\rm Mn}}-2y_{\bar{\rm As}}$ and $\gamma=\frac{n_{h}}{x_{{\rm Mn}}}$.
}
\end{figure}
\par

We plot in Fig. 2, the measured  $T^{exp}_{C}$ as a 
function of $\gamma=n_{h}/x_{{\rm Mn}}$. In contrast to calculated values,  $T^{exp}_{C}$ is more sensitive to the carrier density, varying  linearly with $\gamma$.
The other important difference with the curves in Fig.1 is that ferromagnetism is still 
observed for rather small values of $\gamma$. Thus, if we assume that As anti-sites  dominate compensation, theory and  experiment would disagree.
As our approach was successful for uncompensated samples, and consistent
with Monte-Carlo\cite{Joseph-MC} we conclude that  As anti-sites do {\it not} dominate compensation. 
% Our semi-analytical approach was successful, and consistent with Monte-Carlo simulations \cite{Joseph-MC}, in reproducing the experimentally observed variation of $T_{C}$ as a function of $x_{{\rm Mn}}$ in uncompensated samples ('optimally annealed'). We argue that the disagreement is because As anti-sites do {\it not} dominate compensation.
\begin{figure}[tbp]
\centerline{
\psfig{file=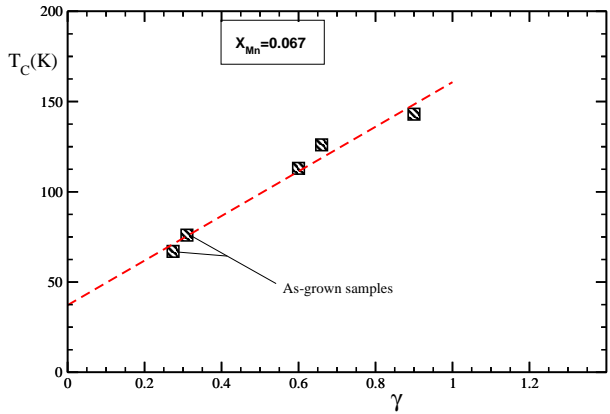,width=7cm,angle=-90}}
\caption{Experimental values of T$_C$ (\cite{Edmonds}) as a function of the measured hole density per magnetic impurity $\gamma=\frac{n_{h}}{x_{{\rm Mn}}}$ for ${\rm Ga}_{1-x} {\rm Mn}_{x}{\rm As}$ at nominal concentration 6.7\%.}
\end{figure}
\par
\begin{figure}[tbp]
\centerline{
\epsfig{file=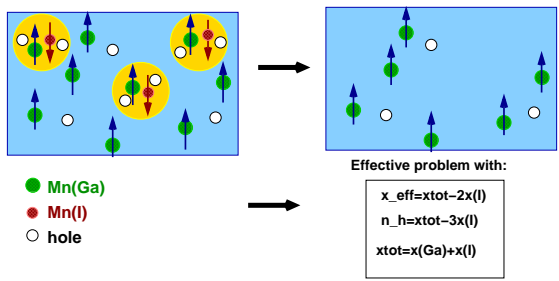,width=7cm,angle=-90}}
%\vspace{-3cm}
\caption{Left side: the up (resp. down) arrows indicate spin of Mn(Ga) (resp. Mn$_{I}$).
The small circles are itinerant carrier (holes). Mn$_{I}$ are double donor and strongly coupled antiferromagnetically to Mn$_{{\rm Ga}}$. Right side: effective model with $x_{\rm eff}$  Mn$_{{\rm Ga}}$ impurities and $n_h$ holes.}
\end{figure}
\begin{figure}[tbp]
\centerline{
\psfig{file= 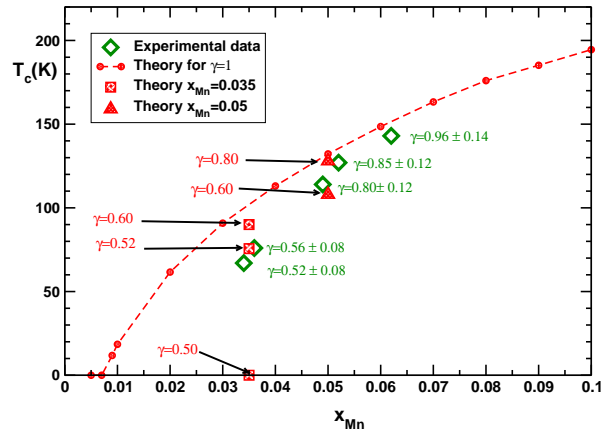,width=7cm,angle=-90}}
\caption{T$_C$  for Mn$_{x}$Ga$_{1-x}$As. The experimental data (diamonds)
are plotted assuming for each sample $x_{{\rm Mn}}=x_{\rm eff}$  and $\gamma =\gamma_{\rm eff}$ (see text).
The values of $\gamma_{\rm eff}$ are shown in the figure.
The small circles , the square and triangles are calculated Curie temperature.
The small circles (dashed line) correspond to uncompensated samples $\gamma=1$  and the square and triangle correspond to $x_{{\rm Mn}}=0.035$ and 0.05 respectively. For these cases, the value of $\gamma$ is shown in the figure.
}
\end{figure}
As already mentioned,  the saturated magnetization\cite{GallagherEdmonds}  at low temperature  indicates that the compensating defects affect both the density of carriers and the density of magnetically active Mn impurities.
Let us now take the alternate limit in which compensation is taken to be 
{\it entirely} due to the presence of Mn interstitial defects Mn$_{I}$.
We denote by $x_{{\rm Mn}}$ , $x_{{\rm Mn}_{{\rm Ga}}}$ and $x_{\rm Mn}({\rm I})$  respectively, the total 
density of Mn, the density of Mn on Ga sublattice and on interstitial location respectively. 
Recent first principle calculations \cite{Masek} and channeling Rutherford backscattering experiments \cite{Yuetal} indicates that Mn interstitials are preferably attracted by Mn$_{{\rm Ga}}$ and tend to form 
pairs of spins with strongly antiferromagnetic couplings. Thus, we suppose 
that Mn$_{I}$ are not completely random, but are only in positions with a Mn$_{{\rm Ga}}$ as a nearest neighbour (see fig. 3).
In writing an effective Hamiltonian, we will eliminate the 
strongly antiferromagnetically coupled pairs of the  Mn$_{I}$ and adjacent  Mn$_{{\rm Ga}}$. They can be assumed, within high   
precision, to  form bound singlet pairs whose effect on the magnetically active
ions is small\cite{BhattLee}. The remaining ``active''  Mn of density $x_{\rm eff}=x_{{\rm Mn}}-2 x_{\rm Mn}({\rm I})$
which are {\it not} directly coupled to a Mn$_{I}$, 
and with the measured carrier density $n_{h}$,  
interact via an effective   Heisenberg model 
with couplings determined by the measured carrier density $n_{h}$.
We make the same calculation
as before but with the effective concentration $x_{\rm eff}$ and compensation $\gamma_{eff}$.
%Because the Mn$_{{\rm Ga}}$ adjacent to interstitials are strongly antiferromagnetically  coupled to Mn$_{I}$, 
%the problem 
%is now mapped to an effective Heisenberg problem, where the density of active Mn is $x_{eff}$ and the density of carrier $n_{h}$ (measured).
Since each Mn$_{I}$ is a double donor, the total density
of carriers is $n_{h}=x_{{\rm Mn}({\rm Ga})} - 2x_{\rm Mn}({\rm I}) =x_{{\rm Mn}}-3x_{\rm Mn}({\rm I})$. 
Thus from each measured  $n_{h}$  we can deduce
the density of Mn$_{I}$ and of {\ }`unpaired' Mn which are, respectively,
$x_{\rm Mn}({\rm I})={\frac{1}{3}}(x_{{\rm Mn}}-n_{h})$ and $x_{\rm eff}={\frac{1}{3}}(x_{{\rm Mn}}+2n_{h})$. 
We also define the effective $\gamma$ parameter as $\gamma_{\rm eff}= \frac{n_{h}}{x_{\rm eff}}=\frac{3 n_{h}}{x_
{\rm Mn}+2n_{h}}$.

In  Fig.4 we show both the experimental data and calculated Curie temperature for various $\gamma$ as a function of $x_{{\rm Mn}}=x_{\rm eff}$. The variable
$x_{\rm eff}$ are  calculated from the values given in \cite{Edmonds} and  the corresponding values of $\gamma_{\rm eff}$ is noted on the figure for each sample. We take the $x_{{\rm Mn}}$ to be the nominal concentration of
each sample; thus we neglect effects of surface inhomogeneities\cite{Wolosetal,GallagherEdmonds,Kirby} in calculating
the {\it bulk} ordering temperature.
First, we observe that the well annealed samples (of highest $T_C$) are in excellent agreement with the calculated values for uncompensated samples, $\gamma=1$ curve ('optimal curve').
We  remark that this optimal curve, which depends on exchange integrals recalculated for 
each concentration,
can nevertheless be well parameterized by the simple form,
up to $x_{\rm Mn}=0.10$,
$ T_{C}\approx A (x_{\rm Mn}-x_c)^{1/2}$ where $x_c=0.0088$, $A=649 ~K $ for Ga(Mn)As.
The samples corresponding to intermediate $T_C$ are still in reasonable agreement with the optimal curve for uncompensated samples at  the {\it effective}
concentration of magnetically active ions. 
The deviation from this optimal line is small but increasingly visible for as-grown samples.
In order to refine  our calculations we have taken into account that for these samples 
$\gamma_{\rm eff}$ is substantially less than 1.
We therefore performed additional calculations for fixed $x_{{\rm Mn}}=0.035$ and 0.05 and various hole densities.
As in Fig. 1, we vary $\gamma_{\rm eff}$ in this effective Hamiltonian with the addition of anti-sites, here used as a purely calculational device to change the carrier density, whilst keeping the calculation fully self-consistent. 
We now observe that the agreement with the experimental measurements is very good for all the measured samples (as-grown and annealed). For example the as-grown sample which corresponds to $x_{\rm eff} \approx 0.035$ and $\gamma_{\rm eff} \approx 0.52$ agrees very well with the calculated value (square symbol) for the same parameters.
Note that using the above relations  we find that the density of Mn$_{I}$ in as-grown samples
is $x_{\rm Mn}({\rm I}) \approx 0.016$ which corresponds to approximately $25 \%$ of the total Mn density.
This is in very good agreement with the value estimated in ref.\cite{Mahieu} (see Fig.4 of ref.\cite{Mahieu}).

\par

In conclusion,
we have shown that  experimental measurements in samples with fixed nominal magnetic impurity concentration could not be explained assuming As anti-sites as the dominant mechanism
for compensation. On the other hand, with the  assumption  that Mn interstitials dominate, we  obtained an excellent quantitative agreement with the measured $T_C$ in both as-grown and annealed samples.
It may be possible, in varying
sample preparation, to increase the number of anti-sites, 
this will have  a weak effect on the T$_C$, provided
the $\gamma_{eff}$ remains above the region of instability, as seen in 
Figure 1.
For Ga(Mn)As samples, we can write an explicit  first approximation ($\gamma_{\rm eff}=1$) 
using the empirical form  for the optimal curve, by replacing  $x_{\rm Mn}$ by $x_{\rm eff}$:
$ T_C \approx A({\frac {x_{\rm Mn} + 2n_{h} } {3} } - 0.0088)^{1/2}$  where $A=649 ~K$.
To take into account the smaller corrections due to the value of $\gamma_{\rm eff}$
we do not have an explicit analytical form, but numerical corrections can be predicted as in Figure 1.
Hence our combined ab-initio/local-RPA approach is a very powerful tool to study  ferromagnetism in diluted ferromagnetic systems even in the presence of compensating defects. The same approach can be applied to 
macroscopic inhomogeneities, for example surface effects,  which may be necessary to understand thin films and devices.

\par
We would like to thank Dr. K. Edmonds for providing unpublished additional
data concerning measured critical temperatures of (GaMn)As. We are grateful to O. Cepas and E. Kats for his comments and carefully reading the manuscript. We also thank  B. Barbara, R. Bouzerar, J. Cibert and C. Lacroix for interesting and fruitful discussions. JK acknowledges the financial support from the Grant agency of the Academy of Sciences of Czech Republic (A1010203) and the Czech Science Foundation 202/04/0583.


\begin{references}
\bibitem{Ohno}H. Ohno, Science {\bf 281},951 (1998).
\bibitem{Edmonds}K. W. Edmonds et al, Phys. Rev. Lett. {\bf 92},037201 (2004),
K. W. Edmonds K.Y. Wang,R.P. Campion, B.L. Gallagher, C.T. Foxon, Appl. Phys. Lett. {\bf 81},4991 (2002). Additional values of T$_c$ were provided by Edmonds et al ( private communication).
\bibitem{Masek} J. Ma\v{s}ek and F. M\'aca Phys. Rev. B 69, 165212 (2004)
\bibitem{Wolosetal} A. Wolos, M. Kaminska, M. Palczewska, A. Twardowski, X. Liu, T. Wojtowicz and J.K. Furdyna
 Journal of Applied Physics, {\bf 96}, 530 (2004).
\bibitem{GallagherEdmonds} 
Wang KY, Edmonds KW, Campion RP, Gallagher BL, Farley NRS, Foxon CT, Sawicki M, Boguslawski P, Dietl T
 Journal of Applied Physics, {\bf 95},  6512-6514 (2004). 
\bibitem{Kirby} B.J. Kirby, J.A. Borchers, J.J. Rhyne, S.G.E. te Velthuis, A. Hoffmann, K.V. O'Donovan, T. Wojtowicz, X. Liu, W.L. Lim and J.K. Furdyna Phys. Rev. B. {\bf 69}, 081307 (2004),
B.J. Kirby, J.A. Borchers, J.J. Rhyne, K.V. O'Donovan, T. Wojtowicz, X. Liu, Z. Ge, S. Shen and J.K. Furdyna Appl. Phys. Lett. {\bf 86}, 072506 (2004).
\bibitem{Glasetal} F. Glas, G. Patriarche, L. Largeau and A. Lema\^itre, Phys. Rev. Lett. {\bf 93}, 086107 (2004).
\bibitem{Tuomistoetal}  F. Tuomisto, K. Pennanen, K. Daarinen and J. Sadowski, 
Phys. Rev. Lett. {\bf 93}, 055505 (2004).

\bibitem{Bouzerar-epl} G. Bouzerar, T. Ziman and J. Kurdnovsk\`y, European Physics Letters, {\bf 69},  812-818 (2005).

\bibitem{Josephetal}J. Kudrnovsk\`y, I. Turek, V. Drchal, F. Maca, P. Weinberger, P. Bruno  Phys. Rev. B, {\bf 69}, 115208 (2004).
\bibitem{Georges2} G. Bouzerar and P. Bruno, Phys. Rev. B. {\bf 66 }, 0114410 (2002) 
\bibitem{Bouzerar-apl} G. Bouzerar, T. Ziman and J. Kudrnovsk\`y, Appl. Phys. Lett. {\bf 85} 4941 (2004).
\bibitem{Wessels} Y.L. Soo, S. Kim, Y.H. Kao, A.J. Blattner, B. Wessels, S. Khalid, C. Sanchez Hanke and
C.C. Kao ,  Appl. Phys. Lett. {\bf 84} 481 (2004); A.J. Blattner, P.L. Prabhumirashi, V.P. Dravid and B.W. Wessels,
J. Crystal Growth {\bf 259}, 8 (2003).

\bibitem{Joseph-MC} L. Bergqvist, O. Eriksson, J. Kudrnovsk\'y, P.A. Korzhavyi, and I. Turek, Phys. Rev. Lett. {\bf 93}, 137202 (2004).  K. Sato, W. Schweika, P.H. Dederichs, and H. Katayama-Yoshida, Phys. Rev. B {\bf 70}, 201202 (2004).




\bibitem{Yuetal} K.M. Yu, W. Walukiewicz, T. Wojtowicz, I. Kuryliszyn, X. Liu, Y. Sasaki and J.K. Furdyna
Phys. Rev. B {\bf 65}, 201303(R) (2002).
\bibitem{BhattLee}R. N. Bhatt and P. A. Lee, Phys. Rev. Lett. {\bf 48}, 344 (1982)
\bibitem{Mahieu} G. Mahieu, P. Condette, B. Grandidier, J.P. Nys, G. Allan, D. Sti\'evenard, Ph. Ebert, H. Shimizu and 
 M. Tanaka Appl. Phys. Lett. {\bf 82} 712 (2003).


\end{references}
\end{document}